\documentclass[a4paper, 11pt]{article}
\pdfoutput=1
\usepackage{graphicx}
\begin{document}
\title{Modelling time and vintage variability in retail credit portfolios: the decomposition approach}
\author
{
    Jonathan J. Forster, Agus Sudjianto\\ \\
Analytics and Modelling, Risk Division, Lloyds Banking Group\thanks{This article represents the views and analysis of the authors and should not be taken to represent those of Lloyds Banking Group.}\\
Email:~{\tt \{Jonathan.Forster\}\{AgusSudjianto\}@Lloydsbanking.com}
 }
\maketitle

\begin{abstract}
 In this paper, we consider the problem of modelling historical data on retail credit portfolio performance, with a view to forecasting future performance, and facilitating strategic decision making. We consider a situation, common in practice, where accounts with common origination date (typically month) are aggregated into a single {\it vintage} for analysis, and the data for analysis consists of a time series of a univariate portfolio performance variable (for example, the  proportion of defaulting accounts) for each vintage over successive time periods since origination.  
An invaluable management tool for understanding portfolio behaviour can be obtained by decomposing the data series nonparametrically into components of exogenous variability (E), maturity (time since origination; M) and vintage (V), referred to as an EMV model. 
For example, identification of a good macroeconomic model is the key to effective forecasting, particularly in applications such as stress testing, and identification of this can be facilitated by investigation of the macroeconomic component of an EMV decomposition.
We show that care needs to be taken with such a decomposition, drawing parallels with the Age-Period-Cohort approach, common in demography, epidemiology and sociology. 
We develop a practical decomposition strategy, and illustrate
our approach using data extracted from a credit card portfolio. 
\end{abstract}

\section{Introduction}
We consider the problem of modelling historical data on credit portfolio performance, with a view to forecasting future performance. Suppose that we have a quantity of interest, $Y$, such as a default rate, which is observed over time for a portfolio of loans; we use loan here as a generic term for the credit product under consideration. We assume that $Y$ is observed over a series of discretely indexed time periods ($t=1,\ldots,T$) which we assume here to be months, and that $Y$ is also separately observed by age ($a=1,\ldots,A$) representing the maturity, or time on books, of the loan. The maximum maturity, $A$, may be fixed (for example where the portfolio represents fixed-term loans) or may simply be the largest value observed in the data.
In addition to time, $t$, and maturity, $a$, the other factor which is initially used to explain variability in $Y$ is the vintage, $v$, of the loan, which indexes the time of origination of a loan. The vintage, $v$, can be derived from $a$ and $t$ by the relationship
\begin{equation}\label{vin}
v=t-a
\end{equation}
under the, essentially arbitrary, choice of labelling which assumes that the vintage which originates in the first time period of observation is labelled 1. In the case where we also observe data for vintages which originate before the first observation time, then these vintages are indexed $0,-1,-2,\ldots$.
The quantity of interest, $Y$, may also be separately observed according to other potentially explanatory account-level variables, but we assume initially that separate analyses are carried out for each segment of the portfolio derived from such variables with no ‘borrowing of strength’  of parameter estimation across segments. Hence, we assume that accounts within a vintage are exchangeable for the purposes of analysis, and that there is no loss of information by aggregating to vintage level. This assumption is plausible for a retail portfolio, but less so for a portfolio of commercial loans.

We can represent the data within a rectangular array with age indexing rows, and time indexing columns as in Table 1.

\begin{center}
\begin{tabular}{c@{\hspace{12mm}}ccccccc}
\hline
&&&&t&&&\\
$a$&1&2&3&$\cdots$&$T-2$&$T-1$&$T$\\
\hline
0&$Y_{01}$&$Y_{02}$&$Y_{03}$&$\cdots$&$Y_{0\,T-2}$&$Y_{0\,T-1}$&$Y_{0\,T}$\\
1&$Y_{11}$&$Y_{12}$&$Y_{13}$&$\cdots$&$Y_{1\,T-2}$&$Y_{1\,T-1}$&$Y_{1\,T}$\\
$\vdots$&$\vdots$&$\vdots$&$\vdots$&$\ddots$&$\vdots$&$\vdots$&$\vdots$\\
$A-1$&$Y_{A-1\,1}$&$Y_{A-1\,2}$&$Y_{A-1\,3}$&$\cdots$&$Y_{A-1\,T-2}$&$Y_{A-1\,T-1}$&$Y_{A-1\,T}$\\
$A$&$Y_{A1}$&$Y_{A2}$&$Y_{A3}$&$\cdots$&$Y_{A\,T-2}$&$Y_{A\,T-1}$&$Y_{A\,T}$\\
\hline
\end{tabular}

Table 1: Data structure 
\end{center}

Typically, not every element in Table 1 is observed. For example, where the credit product does not have a fixed maturation date, the maximum maturity $A$ observed in the portfolio will typically only be observed at the final time period, so the bottom left hand triangle of data is missing.
Vintages correspond to observations which lie on a common diagonal of Table 1 (as represented in Table 2) some of which which originated before the observation period and hence for which  earlier (before $t=1$) observations are not part of the observed data. Observations which represent a common vintage are on diagonals of Table 1, as represented in Table 2.

\begin{center}
\begin{tabular}{c@{\hspace{12mm}}ccccccc}
\hline
&&&&t&&&\\
$a$&1&2&3&$\cdots$&$T-2$&$T-1$&$T$\\
\hline
0&$v=1$&$v=2$&$v=3$&$$&$v=T-2$&$v=T-1$&$v=T$\\
1&&$v=1$&$v=2$&$\ddots$&&$v=T-2$&$v=T-1$\\
2&&&$v=1$&$\ddots$&&&$v=T-2$\\
$\vdots$&&&&$\ddots$&&&\\
\hline
\end{tabular}

Table 2: Relationship between vintages, age and time (excluding vintages originating before observed data period) 
\end{center}

Vintage analysis refers to a methodology, common in credit portfolio analysis, of analysing the variability of portfolio performance, as a function of maturity, by vintage. Hence, the vintage time series, presented as the diagonals of Table 1 (indexed as presented in Table 2) are analysed to compare how performance as a function of age varies between vintages; see Siarka (2011) and references therein for a more detailed description.
The aim of the decomposition approach, as described in this paper, is to explain the variability in $Y$ as a function of the three factors, Maturity (indexed by $a$), Vintage (indexed by $v=t-a$) and Exogeneous (indexed by $t$), and hence to predict $Y$ into the future. We refer to this approach as EMV modelling. 

The decomposition is attractive as, in principle, it allows three important components of variability in the response to be separated in a way which leads to straightforward interpretation. This provides an invaluable management tool for understanding the performance of a portfolio, to aid strategic decision-making, without requiring detailed analysis at account level. The exogeneous component reflects the sensitivity of the portfolio response to the economic cycle. This is the systematic response of the portfolio to exogeneous variability, with idiosyncratic features of the portfolio removed. The maturity component reflects the credit term structure for the portfolio under consideration, representing the natural pattern of variation of the response over time since origination for the credit product under investigation. Finally, the vintage component reflects aspects shared by accounts with common origination dates. It is the idiosyncratic factor which  accounts for aspects such as changing business practices concerning credit approvals and marketing strategies. Analysis of the vintage component can provide detailed insight into, and understanding of, portfolio behaviour.

If forecasting is a goal, it will be necessary to relate variability of the response to parametric functions of  m and/or t and/or v, or to functions of predictable covariates associated with them, for example macroeconomic variables whose values can be predicted into the future. Initially, however, we consider a 'nonparametric' specification, which assumes no relationship between factor levels, or association with external covariates. This provides an exploratory analysis which should help to identify a more parsimonious (and forward predictable) specification at the next stage. 

Breeden (2007) develops an approach to exactly this kind of nonparametric EMV decomposition, which he calls {\it dual time dynamics} because of the dual time scales (calendar-time and maturity) which are used to index the data in Table 1. This extends the approach beyond standard vintage analysis which typically does not explicitly model variability with calendar-time.
The approach investigated in this paper is a special case of Breeden's dual-time dynamics, which is commonly used in credit risk modelling, admits straightforward fitting and which takes an identical form to Age-Period-Cohort or APC models, common in Demography, Epidemiology and Sociology. Age, Period and Cohort correspond to Maturity, Time (Exogeneous) and Vintage respectively. In those applications the response variable of interest, $Y$, is often numbers of deaths or diagnoses of a particular disease or condition, modelled as a function of a subject's age and of calendar time; see Fienberg and Mason (1985),  Clayton and Schiffers (1987) or Robertson et al (1999) for a description of APC modelling. APC models have been very widely studied and debated and, in this paper, we use this existing literature to inform our strategy for EMV modelling.

Let $Y$ denote the vector with components $Y_{at}$, $a=1,\ldots, A$, $t=1,\ldots ,T$, and let $\theta$, with components $\theta_{at}$ be a corresponding linear predictor. We assume that the linear predictor $\theta$ can be expressed as $\theta=X\beta$
for a model matrix $X$ containing values derived from observable explanatory variables, and an unknown coefficient vector $\beta$, and that we have a linear model
\begin{equation}\label{lm}
	E[g(Y_{at}) ]=\theta_{at}, \qquad a=1,\ldots, A, t=1,\ldots ,T
\end{equation}
for some function $g(Y)$, or a generalized linear model ($g$-linear) for $Y$, in which case
\begin{equation}\label{glm}
	g[E(Y_{at}) ]=\theta_{at}, \qquad a=1,\ldots, A, t=1,\ldots ,T.
\end{equation}
Where $g$ is the logarithm function and $Y$ represents a count of the number of default events in the portfolio, then  $\theta$ may be considered as the (discrete-time) log-hazard function for the default process, where the time-to-default is represented by $a$.

\section{Nonparametric EMV specification}

\subsection{Nonidentifiability and constrained estimation}

We assume a 'nonparametric' additive specification,
\begin{equation}\label{emv}
	\theta_{at}= \beta_0+\beta_a^M+\beta_t^E+\beta_v^V,
\end{equation}
where $\beta_0$ is an intercept, and $\beta^M$, $\beta^E$, and $\beta^V$ are parameter vectors for unstructured effects  of maturity, time and vintage respectively. In terms of the two-way layout in Table 1, this model may be thought of as an additive model plus a structured maturity-time interaction, parameterised by $\beta^V$. Similarly, in the vintage analysis framework, where $a$ and $v$ are the defining indices, we again have an additive model, but now with a structured maturity-vintage interaction, parameterised by $\beta^E$. For data on default events, where $\theta_{at}$ can be interpreted as a log-hazard function, then the additive specification (\ref{emv}) represents a proportional hazards model, with baseline log-hazard $\beta^M$, and with $\exp \beta^E$ and $\exp \beta^V$ representing  hazard multiplier effects corresponding to calendar-time and vintage respectively. We describe (\ref{emv}) as nonparametric, because the length of each of the parameter vectors $\beta^M$, $\beta^E$, and $\beta^V$ depends on the size of the observed data rather than being fixed as a function of the model.

Breeden (2007; equation (19)) also considers models of exactly this form, and as described in Section 2.1, they are exactly analagous to APC models used in other fields. As with any standard factorial model, there is a natural identifiability conflict between a factor effect and the scalar intercept ($\beta_0$), which is typically addressed by a single linear constraint on the factor effects (or at least on their estimators). Here, we assume, without loss of generality, the zero-mean constraints
\begin{equation}\label{ident}
\sum_{a} \beta_a^M=\sum_{v} \beta_v^V=\sum_{t} \beta_t^E=0
\end{equation}
However, the relationship between $a$, $t$ and $v$ in (\ref{vin}) also imposes a further identifiability conflict, in that, because $v+a-t=0$, then, for any $\delta$, 
\begin{eqnarray}
\beta_0+\beta_a^M+\beta_t^E+\beta_v^V
&=&\beta_0+\beta_a^M+\beta_t^E+\beta_v^V+ \delta(v+a-t)\nonumber\\ 
&=&\beta_0+(\beta_a^M+\delta a)+(\beta_t^E-\delta t)+(\beta_v^V+ \delta v)\label{ni}
\end{eqnarray}

Hence, a linear drift in $\theta$ over time cannot, on the basis of data analysis alone, be attributed uniquely as a time effect, a maturity effect or a vintage effect, or indeed any given combination of these. This feature has long been recognised in the sociology and epidemiology literature on APC models, where there has been a vigorous debate over its significance, and over how to deal with it in practice. See, for example, Glenn (1976), Goldsetin (1979), Kupper et al (1983), O'Brien (2011) and Fu et al (2011).
Bosman (2012) provides a useful summary of the APC literature, and its connection with vintage modelling in credit portfolio analysis.
What is undisputed, though, is that any estimable function, $\phi=l^T \beta$ will have a unique unbiased estimator, independent of any single linear constraint that might be imposed to overcome the identifiability conflict inherent in (\ref{ni}).
Furthermore, it is important not to interpret such an identifiability constraint as a model constraint. Whichever constraint is applied, the model for the observed data, and the quality of fit of the model to the observed data is identical. Nothing in the observed data can inform us about the relative appropriateness of two particular constraints in a model of the form of (\ref{emv}). 
Where the constraint matters, however, is in forecasting, as different implied constraints for the data-generating parameters can have different implications for how those parameters are forecast beyond the range of the observed data.

Note that the identifiability conflict implied by (\ref{ni}) can also arise in simpler models than the full nonparametric specification implied by (\ref{emv}). The simplest such model is one where  $e$, $m$ and $v$ are included in the model directly as quantitative covariates; see Tu et al (2011) for an example. Between the two extremes of linear dependence on $e$, $m$ and $v$ and the completely unstructured factor effects of (\ref{emv}), is a generalised additive model (GAM) where one or more of the E, M and V effects are modelled as smooth functions in $e$, $m$ and $v$. Because the class of smooth regression functions subsumes the   simple linear model, the identifiability conflict arises there too. Breeden (2007, Section 3.6) provides a good example of this, where the results of fitting a GAM are found not to closely match the E, M and V functions which were used to generate a set of test data. Although not discussed explicitly by Breeden (2007) this behaviour is largely due to the identifiability conflict, as it can be seen that much closer resolution with the generating model can be achieved by addition of the same linear function to the estimated M and V effects, with a balancing subtraction from the E effect, exactly as implied by (\ref{ni}).

Identifiability can be imposed by any single linear constraint on the underlying model parameters. For example, one might impose the constraint that the final two vintage effects are the same, so that
\begin{equation}\label{SASc}
\beta_T^V=\beta_{T-1}^V.
\end{equation}
Yang et al (2004) describe the “intrinsic estimator”, defined as $c^T \beta=0$,
where $c$ is the null vector of the model matrix $X$ (i.e. $Xc=0$) arising as a result of the relationship in (\ref{vin}). Therefore, if $\beta$ is arranged as
$\beta=(\beta_0,\beta_a^M,\beta_t^E,\beta_v^V )$,
we have
\begin{equation}\label{c}
c=(0,1,\ldots,A,-1,\ldots,-T,1,\ldots,T)^T,
\end{equation}
assuming no historic vintages in the data. The intrinsic constraint can be thought of as estimating $\beta^{\rm int}$, the uniquely determined projection of $\beta$ onto the orthogonal complement of $c$ (or equivalently the column space of $X^T X$).  Hence, for any equivalent $\beta$, we have
$\beta^{\rm int}=(I-\frac{1}{c^T c} cc^T )\beta$,
and so $\beta^{\rm int}$ is the parameter of minimum length amongst all equivalent $\beta$. 
A similar, implicit, constraint (but based on an orthonormalised model matrix) is imposed when the model is fit using Partial Least Squares (see Wold, 1985), as suggested, for the linear APC model, by Tu et al (2011).
Yang et al (2004)  argue that, in the absence of any external justifications for imposing a particular constraint, the intrinsic estimator has some appealing properties, but we argue against setting a single default constraint or implicitly imposing a constraint by using a particular fitting strategy such as PLS, when using an EMV decomposition, as we do here, to aim to identify a possible forecasting model. 

For a linear model, any estimator of the parameter vector $\beta$ under a single linear constraint can be expressed as
\begin{equation}\label{gls}
	\hat{\beta}=(X^T X)^- X^T g(Y)	
\end{equation}
where $(X^T X)^-$ is a particular generalized inverse of $X^T X$. Furthermore, for any linear or generalised linear model, any constrained estimator can be obtained from any other by linear reparameterisation, 
\begin{equation}\label{trans}
	\hat{\beta}^{(2)}=(X^T X)^- X^T X\hat{\beta}^{(1)}
\end{equation}
where $(X^T X)^-$ is the particular generalized inverse of $X^T X$ which gives the estimator, $\hat{\beta}^{(2)}$, under the new constraints.

\subsection{Practical nonparametric estimation}

When fitting linear or generalised linear statistical models, strict nonidentifiability rarely causes a problem (unlike, say high multicollinearity, which can be characterized as ‘close to non-identifiable’). Most software packages (for example SAS and R) recognise that $X$ is not of full rank, and provide {\it a} solution to the relevant estimating equations, thereby implicitly applying a parameter constraint. For linear models, this means
providing a solution to the normal equations
\begin{equation}\label{ne}
X^T X\hat{\beta}=X^T g(Y)
\end{equation}
in the form of (\ref{gls}) for a generalized inverse $(X^T X)^-$ which can be thought of as providing an estimate for $\beta$ under a particular constraint. SAS (PROC GLM or PROC GENMOD) effectively imposes the constraint (\ref{SASc}), while R (under its default treatment of the standard nonidentifiability for factorial models) imposes the constraint
\begin{equation}\label{Rc}
	\beta_T^V=\beta_1^V.
\end{equation}
In both cases, this assumes that the vintage effect is the last to be entered into the model formula.

As discussed in Section 2.1, the constraint used in fitting is simply a convenient device to obtain a unique estimator, which for a linear model is given by a solution of the normal equations (\ref{ne}), corresponding to a particular generalised inverse of $X^T X$.
Any $\hat{\beta}$ can be transformed after fitting to a more substantively meaningful solution, assuming that one exists, by the linear transform (\ref{trans}), where $(X^T X)^-$ is a generalized inverse of $X^T X$ corresponding to the required constraints. Equivalently, and practically more straightforward, any two constrained estimators are related by
\begin{equation}\label{diff}
\hat{\beta}^{(2)}=\hat{\beta}^{(1)}+c\gamma
\end{equation}
so differ by a scalar ($\gamma$) multiple of the vector $c$, given in (\ref{c}). Hence, if $d^T \beta=0$ is the constraint which we require to be satisfied by the new estimator, it can be obtained using (\ref{diff}), where $\gamma$ solves
\begin{equation}\label{consol}	
0=d^T \hat{\beta}^{(1)}+d^T c\gamma.
\end{equation}
More generally, when there are multiple sources of nonidentifiability, defined by $XC=$0 for some matrix $C$, then constrained estimators under the constraints $D^T \hat{\beta}=0$ can be obtained by replacing $c$ and $d$ in (\ref{diff}) and (\ref{consol}) by $C$ and $D$. This applies equally well to generalised linear models (\ref{glm}).

\subsection{Meaningful constrained estimation}

Our main motivation for fitting a nonparametric EMV model is for exploratory analysis to understand the components of the variability in the response, prior to building a forecasting model based on, for example, macroeconomic covariates. It is therefore important that the constraints under which the estimates are presented do not disguise important variation which has the potential to be modelled effectively.  In the EMV structure, the nonidentifiability implied by ({\ref{ni}) therefore forces us to consider how to meaningfully attribute any time drift in the underlying linear predictor, rather than rely on a default or implicit solution.

The final model fitted aims to explain time (exogeneous) variability through macroeconomic covariates and maturity variability as some smooth function of  age. Typically, vintage is not explicitly modelled, remaining in the model as a ‘nonparametric’ effect. For the purposes of forecasting, any time drift in the response may be  forecast forward as a result of the combination of projected exogeneous (and maturity) effects with the parameters corresponding to new vintages projected on the basis of being close to current levels, or based on business assumptions about credit decisions going forward. One possibility, therefore, to get a potentially meaningful nonparametric estimate of the exogeneous, maturity and vintage effects, is to constrain the vintage effects, so that they do not grow over time.
Both the SAS (\ref{SASc}) and R (\ref{Rc}) constraints have the effect of constraining the vintage time-drift, defined as a particular pairwise difference, to zero. However, when applied to parameter estimates, both depend on just two parameters, so can be somewhat sensitive to small changes in the data; see Section 3. A global zero-time-drift constraint is to force a regression of vintage effect on vintage number to have slope zero, so that
\begin{equation}\label{vincon}
\sum_{v}(\beta_v^V-\bar{\beta}^V )(v-\bar{v})=0
\end{equation}
which is a linear constraint in $\beta^V$. In practice, it may be desirable that the effects of earlier vintages, which are arguably less relevant when considering how to forecast future vintage effects, are removed from the constraint computation in (\ref{vincon}),  so that the summation and mean computations are then restricted to a set $C_V\subset\{1,\ldots,T\}$. The most recent vintage estimates are based on the fewest months of experience, so caution should also be exercised in  using a set $C_V$ with only a few recent vintages.
In general, we consider that there is a danger in imposing the identifiability constraint on the vintage parameters, in that the between-vintage variability is likely to be the aspect of the portfolio about which {\it a priori} understanding is the weakest.

An alternative to a vintage constraint, is to base the constraint on prior expectation about the variation of the maturity coefficient $\beta^M$ over time. It may be reasonable to expect that, as $a$ increases, the variability in $\beta_a^M$ flattens off, so that $\beta_a^M$ is approximately constant as a function of $a$ for values of $a$ beyond a given threshold. A suitable constraint, based on this consideration, but not explicitly forcing constant maturity beyond threshold, is to force a regression of $\beta_a^M$ on $a$, for $a$ greater than some threshold $A^*$, to have slope zero
\begin{equation}\label{matcon}
\sum_{a>A^*}(\beta_a^M-\bar{\beta}^M )(a-\bar{a})=0
\end{equation}
where $\bar{a}$ is the mean value of $a$ over the region $a>A^*$.

Although 'flat maturity' may seem to be an intuitively plausible constraint, there is a good reason for applying caution in using it indiscriminately. For example, consider a credit card portfolio. It might seem plausible to assume that, once an account has been open for a sufficiently long period, time since origination becomes an irrelevant factor in modelling the probability of  default. This is exactly the consideration which leads to flat maturity. However, it is important to realise that we are modelling data which are aggregated to vintage level. Even if all accounts demonstrate flattening maturity effect, the maturity curve in a model for aggregate data will not typically be flat, if there is heterogeneity, within a vintage, in this limiting maturity effect. In survival analysis, this kind of heterogeneity is called {\it frailty} (see, for example, Therneau and Grambsch, Chapter 9) and it seems certain that some level of frailty will exist when considering accounts within a vintage. In the case where accounts have limiting constant maturity effect (log-hazard) individually, heterogeneity in the limiting hazard means that the accounts with the higher values will be removed from the sample (by defaulting) at a higher rate, with the result that the vintage hazard for the surviving proportion of the vintage will be observed to decrease. This is illustrated in Figure 1, where we display a sample of hazard functions (maturity curves) for a hypothetical vintage each with constant limiting hazard. Also displayed is the resulting vintage hazard function which can be seen to decrease after an initial increase, rather than tend to a constant limit. For a general discussion on the shapes of hazard functions, and the mechanisms by which different hazard functions arise, see Aalen and Gjessing (2001) and accompanying discussion.

\begin{figure}[htb]
\begin{center}
\includegraphics[width=12cm]{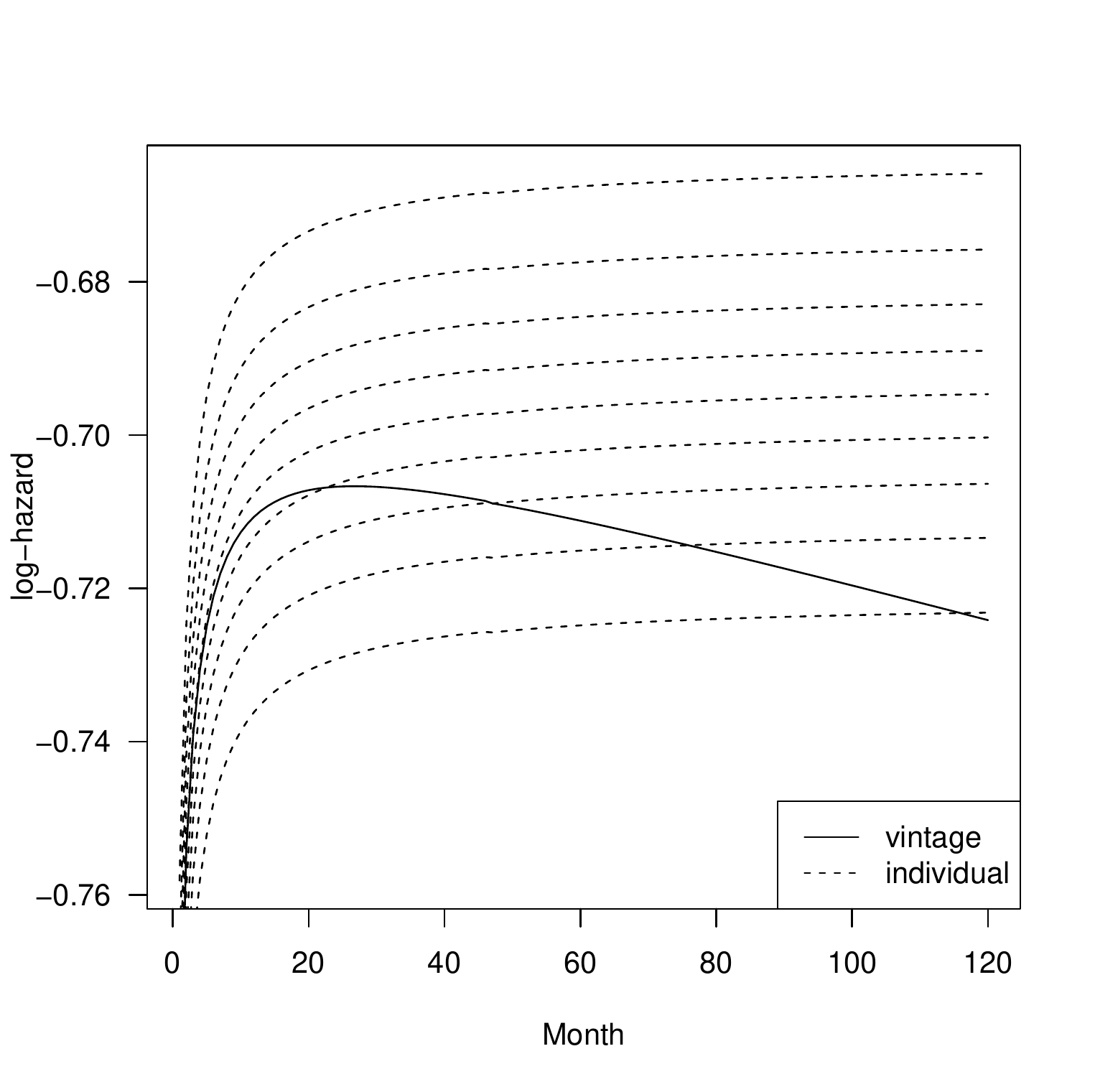}
\caption {This figure illustrates how heterogeneous limiting flat maturity at account level can result in limiting decreasing maturity at vintage level. The dashed lines represent the distribution of account-level maturity curves (they are the curves corresponding to accounts at the $10\%, 20\%,\ldots , 90\%$ quantiles of the frailty distribution). The solid line is the resulting vintage maturity curve. The attenuation in log-hazard at vintage level, resulting from the differentially higher probability of early default of the higher-hazard accounts) is clearly evident.}
\label{frailty}
\end{center}
\end{figure}

\section{Example}

Here, we present an example of the application of the nonparametric EMV decomposition to investigate credit risk data. The data we are using is a segment of a particular credit card portfolio, and the variable we are modelling is the balance moving to a particular arrears grade, as a proportion of total outstanding balance in the portfolio segment. Seven years (84 months) of data are available, from July 2004 to June 2011. Figure \ref{emvrsas} displays the estimated decomposition (\ref{emv}) when the data are analysed using standard linear model fitting routines in R and SAS, with vintage being the final factor input into the model specification, so that the implicit identifiability constraints are (\ref{Rc}) and (\ref{SASc}) respectively.

\begin{figure}[hptb]
\begin{center}
\includegraphics[width=15cm]{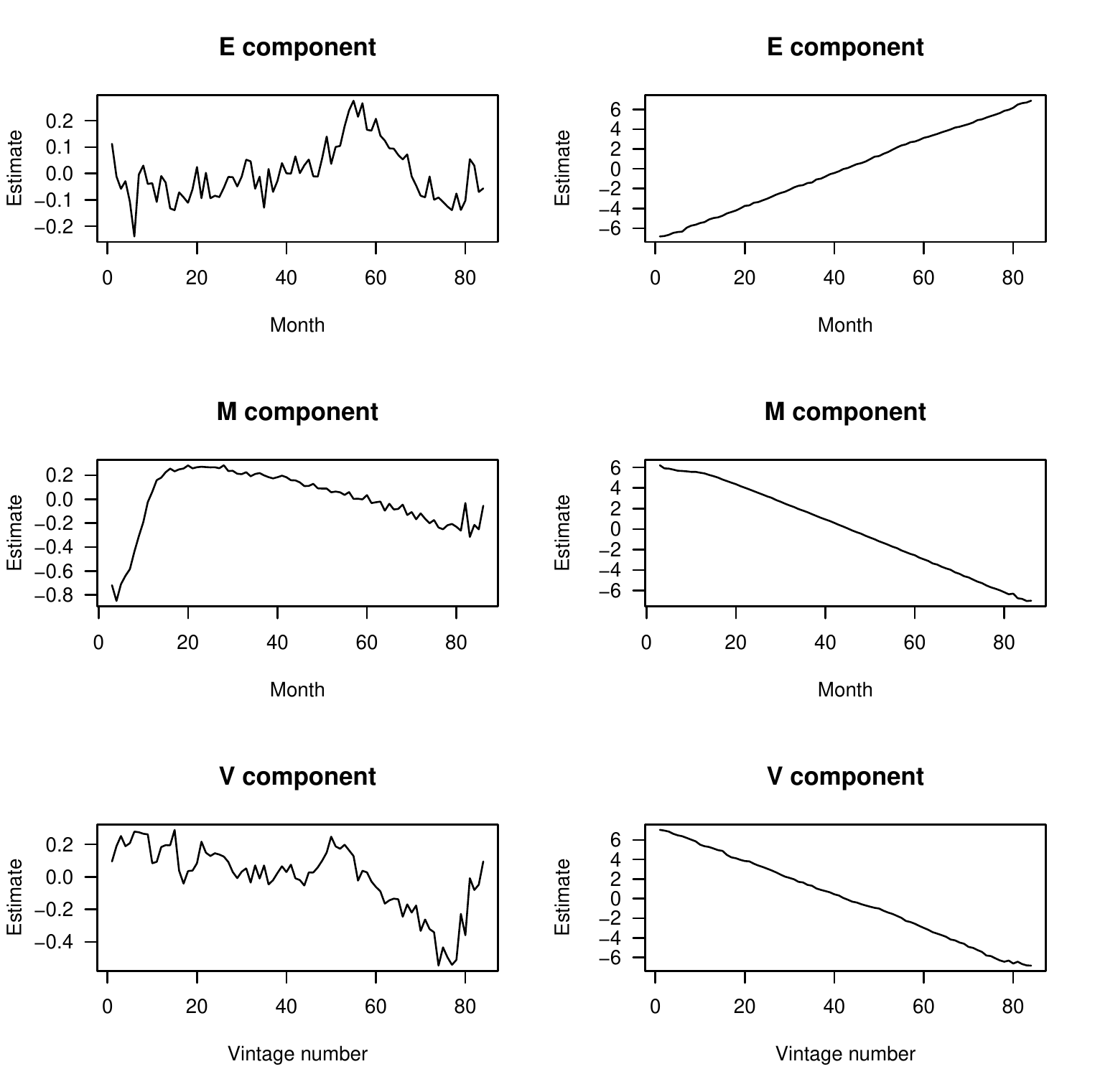}
\caption {EMV decomposition using default fitting and implicit constraints in R (left hand column) and SAS (right hand column)}
\label{emvrsas}
\end{center}
\end{figure}

Although the two sets of estimates in Figure \ref{emvrsas} appear very different, they are exactly equivalent differing only by a common linear drift which cannot be uniquely attributable to E, M or V. In this case the estimates provided by R seem more plausible but that is simply due to good fortune. There is no guarantee that either (\ref{SASc}) or  (\ref{Rc}) provides an appropriate decomposition. However, either of these sets of estimates can straightforwardly be post-processed, using (\ref{diff}) and (\ref{consol}) to obtain estimates consistent with an alternative identifiability constraint. Figure \ref{emvcons} displays the estimated decomposition (\ref{emv}) under three different constraints, representing flat limiting maturity, zero recent vintage trend, and the intrinsic constraint, described in Section~2.1. These three decompositions exhibit greater similarity, although the differences are not negligible. In particular, the exogeneous component appears broadly decreasing under the flat maturity constraint and broadly increasing under the constraint of zero recent vintage trend. Neither of these seem intuitively very plausible, given known macroeconomic variability within the corresponding seven year period.

\begin{figure}[hptb]
\begin{center}
\includegraphics[width=15cm]{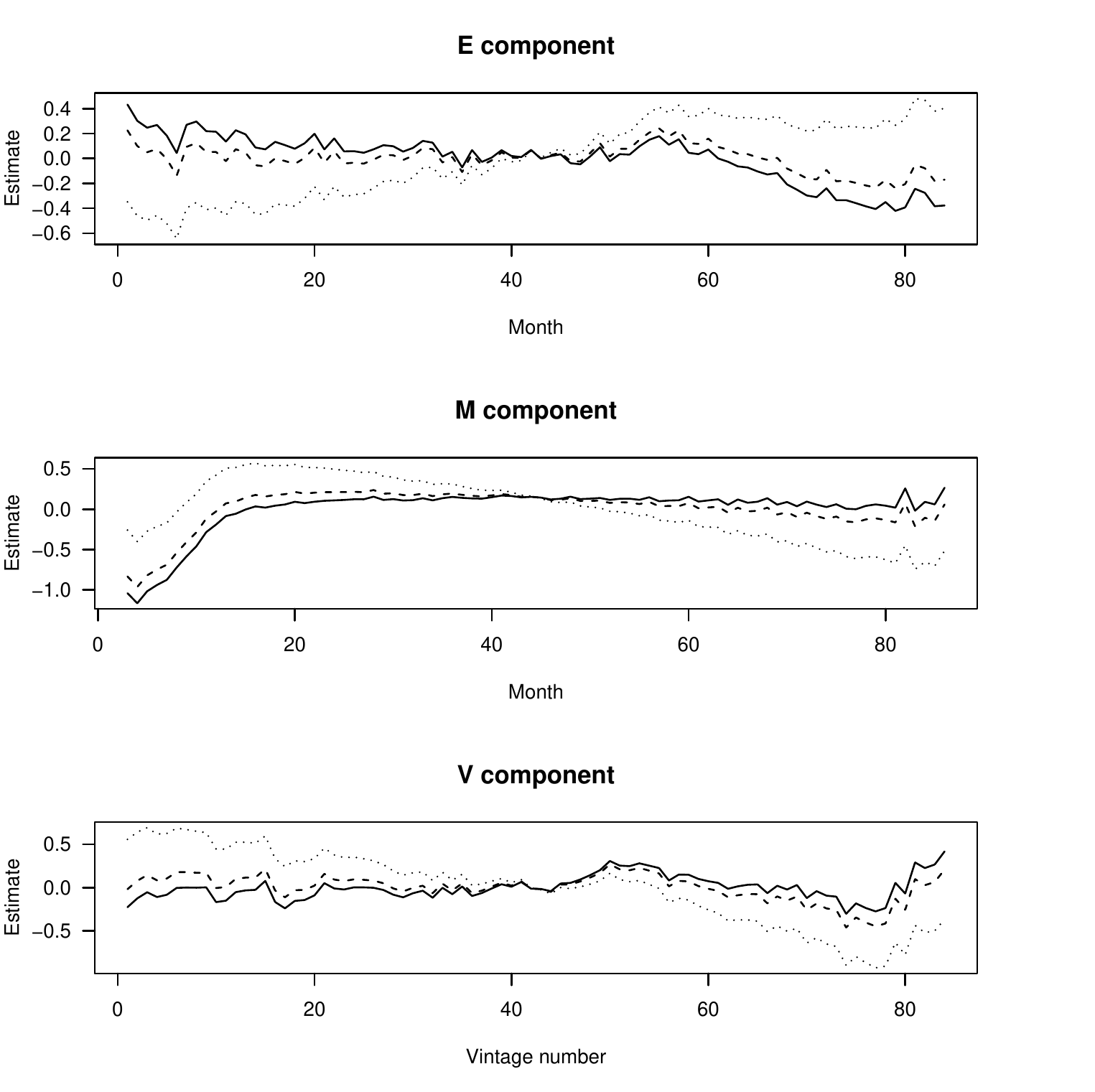}
\caption {EMV decomposition using three different constraints: (i) (solid line) uses (\ref{matcon}) to constrain the average decrease in maturity to be zero after 60 months; (ii) (dashed line) uses the intrinsic constraint, (\ref{diff}) and (\ref{consol}) with $d=c$; (iii) (dotted line) uses (\ref{vincon}) to constrain the vintage trend over the final 18 months to be zero}
\label{emvcons}
\end{center}
\end{figure}

\section{A strategy for nonparametric EMV decomposition}

It is clear that, in identifying an EMV decomposition for further analysis, an element of subjectivity is required in determining a suitable identifiability constraint. Here we describe a possible strategy for credit portfolio decomposition. Both practical experience and intuition suggest that the maturity component exhibits the smoothest variation over time. Therefore, we suggest examining plausible single linear constraints on the maturity curve, as a strategy for comparing equivalent EMV decompositions. For certain portfolio data, such as defaults on a product without a fixed maturation date (such as the credit card data analysed in Section 2.4) there is usually good reason to believe that, at individual account level, sensitivity of response to the age (time-on-books) of the account will diminish over time, leading to a flattening of the maturity curve at account level. As illustrated in Figure~\ref{frailty}, this can lead to a decreasing maturity curve when vintage-aggregated data are analysed. We suggest producing a range of EMV decompositions, based on (\ref{diff}), but using a modified version of (\ref{consol}), where
\begin{equation}\label{consolmat}	
k=d^T \hat{\beta}^{(1)}+d^T c\gamma.
\end{equation}
with the constraint $d^T \beta=0$ derived from
\begin{equation}\label{matcon1}
\frac{\sum_{a>A^*}(\beta_a^M-\bar{\beta}^M )(a-\bar{a})}{\sum_{a>A^*}(a-\bar{a})^2}=k
\end{equation}
and a range of different (negative) values of $k$. This provides a comparison of the limiting flat maturity constraint with a range of alternatives where the maturity decreases over later months, consistent with a model where there is limiting flat maturity at account level, with frailty leading to a decreasing maturity curve in the limit.

Figure \ref{emvmatcons} displays the estimated decomposition (\ref{emv}) using values of $k=0$, $k=-0.01$ and $k=-0.02$. Of these options, the value $k=-0.01$ seems to provide a good balance which allows for some frailty without an excessive level of decrease in the maturity curve in later months. This also has the effect of producing an E-curve which exhibits a plausible response to exogeneous macroeconomic variability. This is illustrated in Figure \ref{emvexog} where the exogeneous series is plotted together with a potential macroeconomic explanatory variable, representing a three-month moving average of increase in UK unemployment. It can be seen that the exogeneous parameter estimates reflect some of the same time characteristics as the macroeconomic covariate, particularly the increase at around month 60, corresponding to a deteriorating macroeconomic climate.

\begin{figure}[hptb]
\begin{center}
\includegraphics[width=15cm]{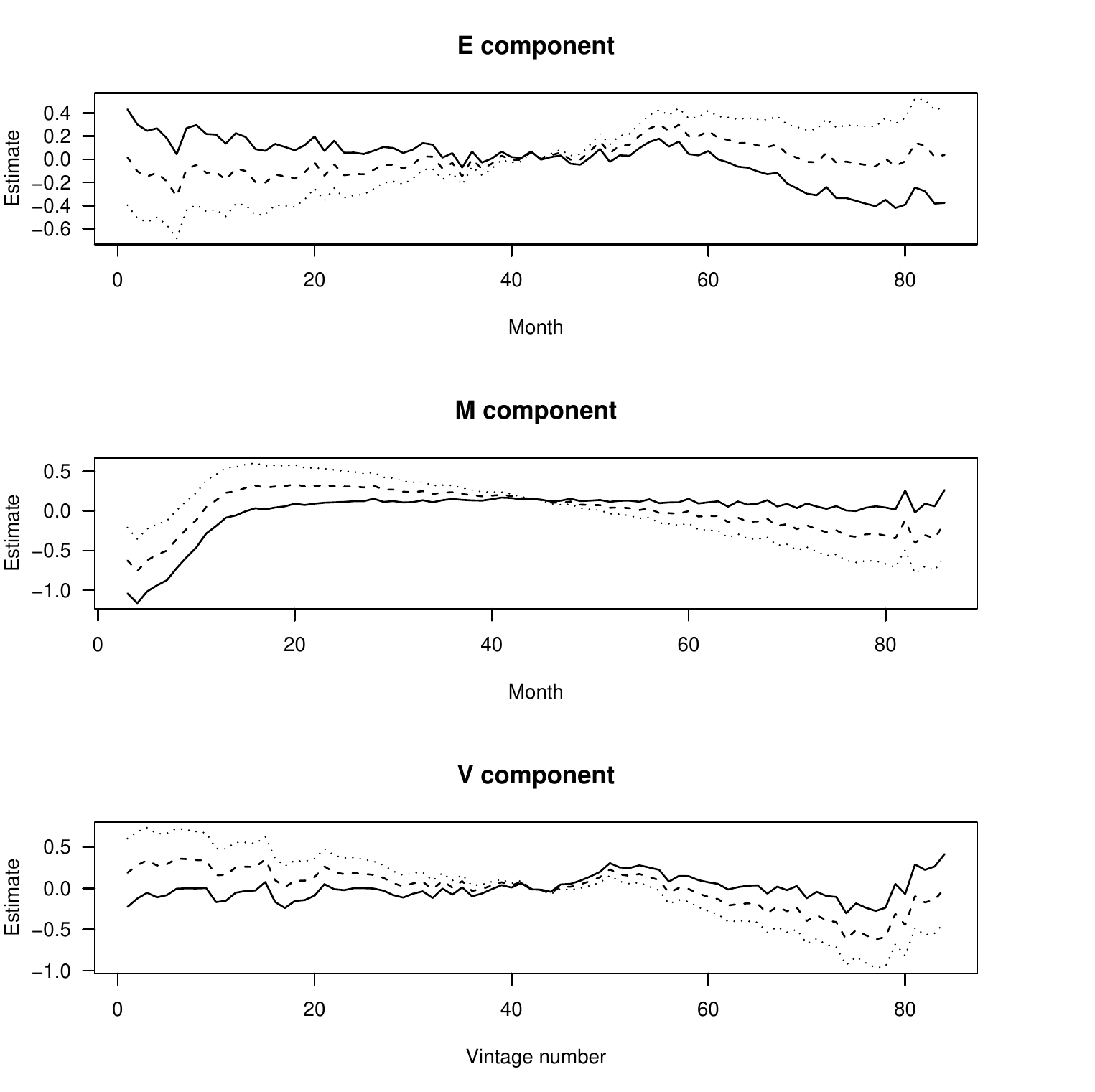}
\caption {EMV decomposition using three different constraints derived from (\ref{matcon1}): (i) (solid line) $k=0$; (ii) (dashed line) $k=-0.01$; (iii) (dotted line) $k=-0.02$}
\label{emvmatcons}
\end{center}
\end{figure}

\begin{figure}[htb]
\begin{center}
\includegraphics[width=12cm]{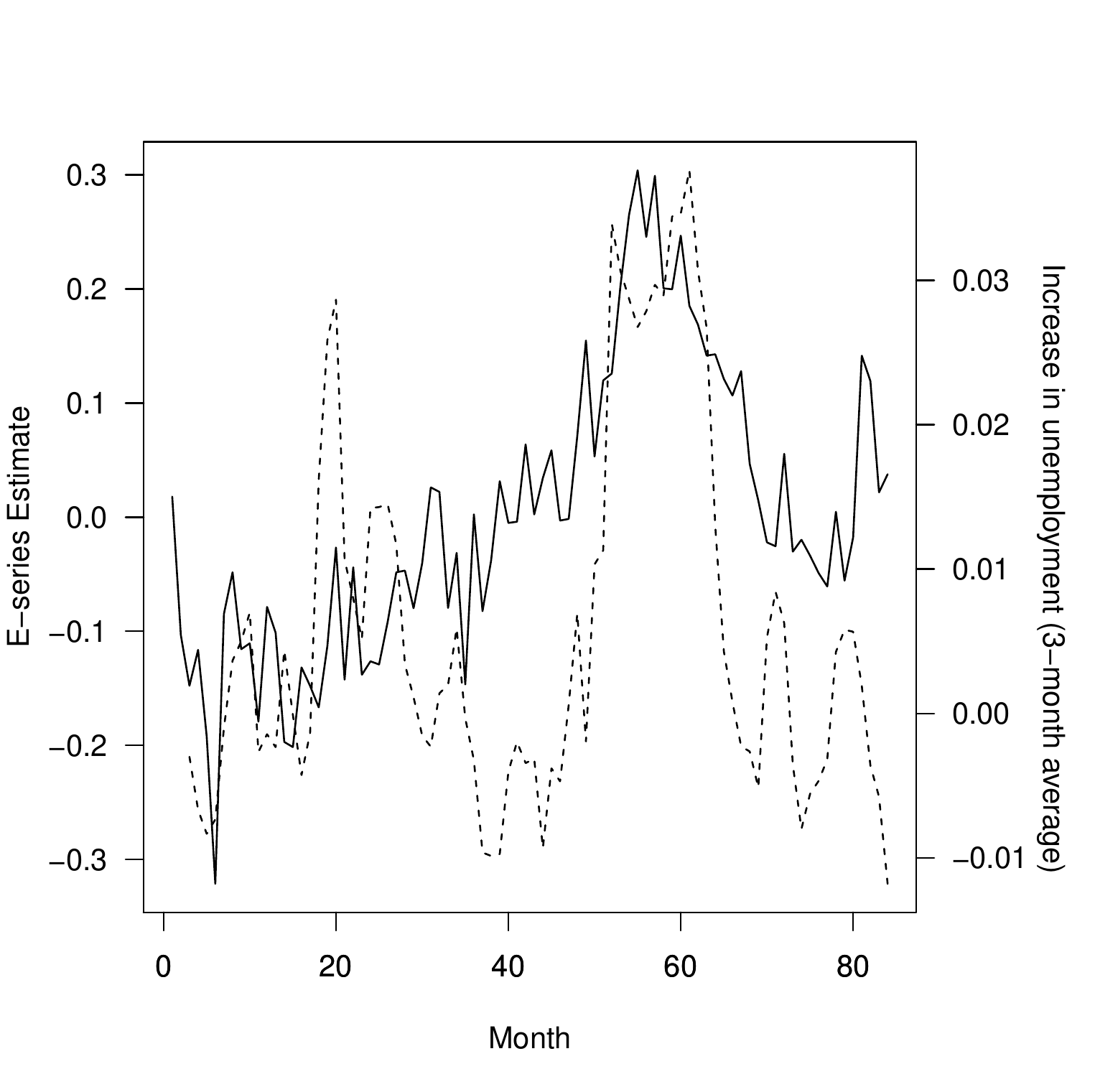}
\caption {Exogeneous component of EMV decomposition (using the constraint derived from (\ref{matcon1}) with $k=-0.01$; solid line) together with  three-month moving average of increase in UK unemployment (dashed line)}
\label{emvexog}
\end{center}
\end{figure}

\section{Semiparametric EMV models}

\subsection{Modelling the exogeneous curve with macroeconomic covariates}

The nonparametric specification described in Section 2 is highly parameterised, with a factor effect for each level of the three factors, Exogeneous, Maturity and Vintage. It is desirable to identify, if possible a more parsimonious model, to reduce the impact of estimation variability on prediction error. Furthermore, as we may require to make forecasts for values of $a$ and $t$ (and hence $v$) outside the range of the observed data, it may be necessary for our forecasting model to explicitly specify a relationship between past and future observations. We focus on replacing the exogeneous parameters $\beta^E$ with a linear predictor based on macroeconomic covariates whose values over the observed data period are known and whose predicted values over the forecast period can be obtained from an external (to the current data) source.
Hence, we adapt the nonparametric additive specification (4), to the (semi-) parametric specification
\begin{equation}\label{par}
\theta_{at}= \beta_0+\beta_a^M+\beta_v^V+\sum_{j=1}^J x_{tj} \beta_j^E ,	
\end{equation}
where $\{x_{tj},t=1,\ldots,T,\;j=1,\ldots,J\}$ are the values of $J$ relevant macroeconomic covariates. Note that the parameter $\beta^E$ now has a different interpretation as a vector of $J$ covariate regression coefficients, rather than a vector of $T$ exogeneous factor effects.
In principle, fitting the model (\ref{par}) is straightforward by standard methods. Indeed, the removal of the exogeneous factor effects means that the identifiability conflict (\ref{ni}) no longer exists in principle and the factor effects $\beta^M$ and $\beta^V$ can be identified by a standard (for example, sum-to-zero) constraint. However, some care is still required, as if the vector $(1,\ldots ,T)$ can be closely predicted by a regression on $\{x_{tj},t=1,\ldots,T,\;j=1,\ldots,J\}$, then a similar identifiability issue can arise in practice. This means that, for example, the inclusion of an exogeneous time trend for modelling portfolios with short  histories will be susceptible to the same identifiability issues as a nonparametric model, with the same care required in presenting parameter estimates.

In order to investigate the fit of a parametric exogeneous model, it may be helpful to compare the predicted time effects implied by the parametric model (the sum on the right hand side of (\ref{par})) with the exogeneous factor effects in a nonparametric model. For this comparison to be meaningful, we need to apply the appropriate nonidentifiability constraint to the nonparametric estimates (the parametric model is not constrained). The correct constraint is one under which the contribution of the exogeneous effects to any time-drift in the response is the same in both models. This is achieved if we ensure that
$$
\sum_{t=1}^T \hat{\beta}_t^{E^*}(t-\bar{t})=\sum_{t=1}^T \hat{\beta}_t^E (t-\bar{t})
$$
where $\bar{t}=(T+1)/2$ and $\hat{\beta}_t^{E^*}=\sum_{j=1}^J x_{tj} \hat{\beta}_j^E$
represents the parametric model estimate of the time effects. This constraint can be achieved by applying (\ref{diff}) with $\hat{\beta}^{(1)}$ replaced by 
$\hat{\beta}^E$, (for any constrained non-parametric estimate), and
$$
\gamma=\frac{\sum_{t=1}^T \hat{\beta}_t^{E^*} (t-\bar{t})-\sum_{t=1}^T \hat{\beta}_t^E (t-\bar{t}) }{\sum_{t=1}^T (t-\bar{t})^2},
$$
the resulting $\hat{\beta}^{(2)}$ being the comparable nonparametric factor effects.

Here we present an example of  a parametric EMV model (\ref{par}) which incorporates the effects of various macroeconomic covariates. The data are the same as were used in Section 2.4 to illustrate the nonparametric model. We use 3 covariates, debt-to-income ratio (DtI), year on year increase in log Unemployment for the UK and year on year increase in log DtI. This model is much more parsimonious than the nonparametric model, as 83 exogeneous dummy variables have been replaced by 3 macroeconomic covariates. To compare the fit of the nonparametric model and this macroeconomic model, we first apply the appropriate identifiability constraint to the nonparametric factor effects, as described above, to obtain Figure \ref{emvpar}.
It is interesting to note that the parametric model, which does not constrain the maturity or vintage effects (other than a standard identifiability constraint, as in (\ref{ident}) to fix their level) results in an estimate for the maturity curve which flattens off at round 18 months thereafter exhibiting a slight decline. This is somewhat satisfying as it seems to accord with intuition about the behaviour of this effect at vintage level. There is no guarantee that this will happen in every example, but if an appropriate exogeneous model is identified, then this estimation approach should identify the correct linear drift in the maturity curve. The shapes of the maturity and vintage curves are largely unaffected by replacing the nonparametric exogeneous curve with a parametric specification.
The overall R$^2$ for the parametric model is 87.9\% (compared with 89.5\% for the nonparametric model), so arguably the parametric model succeeds in identifying the main aspects of variability in the response.

\begin{figure}[hptb]
\begin{center}
\includegraphics[width=15cm]{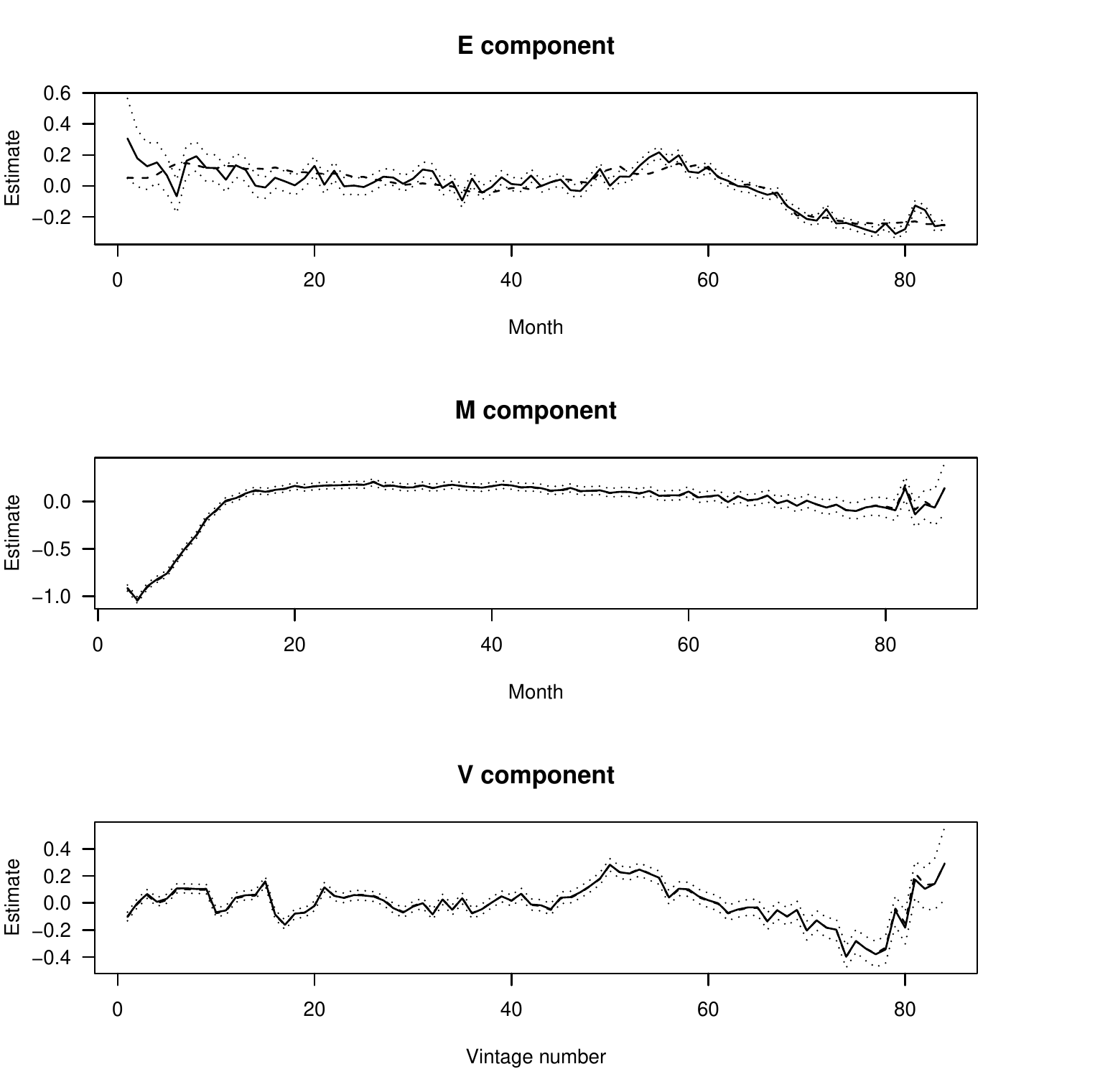}
\caption {Semiparametric EMV decomposition (dashed line) compared with the nonparametric decomposition (solid line) appropriately identified. 95\% confidence intervals for the nonparametric curve are also displayed (dotted lines). }
\label{emvpar}
\end{center}
\end{figure}

\subsection{Modelling the maturity curve}

In the example presented above and other examples of which we have experience, of the three components in the model, it is the maturity factor whose parameters have exhibited the most regular variation as a function of its underlying time index. Hence, it might be possible to replace the nonparametric maturity model by a parametric model, representing a smooth function of age $a$. The appropriate form for such a function will depend on the portfolio under investigation and a detailed investigation is beyond the scope of the current paper.  As described in Section 2.1, when the response represents a conditional default rate (or similar), the EMV model (\ref{emv})  can be thought of as a discrete time proportional hazards model, where the  maturity effect (plus the intercept), $\beta_0+\beta_a^M$ represents log-baseline-hazard. One possible future direction of research is to investigate whether suitable parametric families of maturity curves can be constructed by considering log-hazard functions arising from parametric families used in continuous-time survival analysis, or by the kind of considerations discussed by Aalen and Gjessing (2001). An alternative is a semiparametric approach where the nonparametric estimates are smoothed, for example using a spline regression basis; see Zhang (2009) for details.
Other features such as imposing monotonicity of the maturity curve over certain age ranges might also be incorporated within a semiparametric approach.

In general, a parametric maturity model is not quite so essential for forecasting, as it is only the later maturity values which are unobserved in the data used for estimation and which therefore need to be estimated in order to forecast the response forward in time. For early values of $a$, there is typically a large number of vintages with relevant data, and the unsmoothed, non-parametric, estimate may suffice for forecasting. On the other hand, for larger values of $a$, estimates are based on smaller numbers of vintages and are consequently more variable. For example, the maturity curves in the second panel of Figure \ref{emvpar} become considerably less smooth as the maturity $a$ increases, and the confidence interval widens. A pragmatic approach is to retain a nonparametric approach, but constrain the curve so that all maturity coefficients $\beta_a^M$ where $a$ exceeds a given threshold, $A^*$, are constrained to follow a smooth curve, possibly a straight line. Note that this is a stronger constraint than (\ref{consolmat}) which only requires the maturity curve beyond $A^*$  to be have a particular slope {\it on average} (and hence does not actually correspond to a model assumption, simply a plausible way to resolve the EMV identifiability conflict).

\subsection{Modelling and forecasting vintage effects}

Forecasts of portfolio performance into the future typically require the facility to allow for new vintages with start dates in the future for which the estimation sample contains no data. Currently, common practice is to forecast vintages effects at a common level consistent either with recent experience, or set to reflect expected business decisions concerning portfolio risk. 
An alternative approach is to replace the nonparametric vintage parameters $\beta^V$ with a model based on vintage covariates (if available) for example covariates which summarise the quality of the vintage, or prevailing market conditions, at time of inception. Forecasts are then based on projected values of the covariates (analogous to the way in which exogeneous variation is projected forward).

Vintage quality is a panel-level effect which, in many other application areas, would be incorporated as a random effect. That is, the values of $\beta_v^V$ would themselves be modelled as realisations from a stochastic process. Random effects models are extremely useful in that they allow effects to be predicted for vintages which are unobserved in the data used for estimation, allow the uncertainty of such a prediction to be computed, and automatically calibrate the appropriate level of shrinkage. That is, where the differences between vintages are observed to be small, random effects estimation allows considerable borrowing of strength across vintages, resulting in more efficient prediction.
Most random effects models assume that the random effects are independent and identically normally distributed with zero mean. For EMV models, the assumption of independence may not be realistic, as we may expect vintages to be serially correlated. Hence, a model for $\beta_v^V$ may  be a time series model. There is a precedent for this kind of model in that APC models for human mortality forecasting often include a cohort (vintage) effect which is assumed to follow some kind of ARIMA process. Figure \ref{emvre} displays the estimated vintage effects under two different random effects specifications for the nonparametric analysis of Section~3.

\begin{figure}[htb]
\begin{center}
\includegraphics[width=12cm]{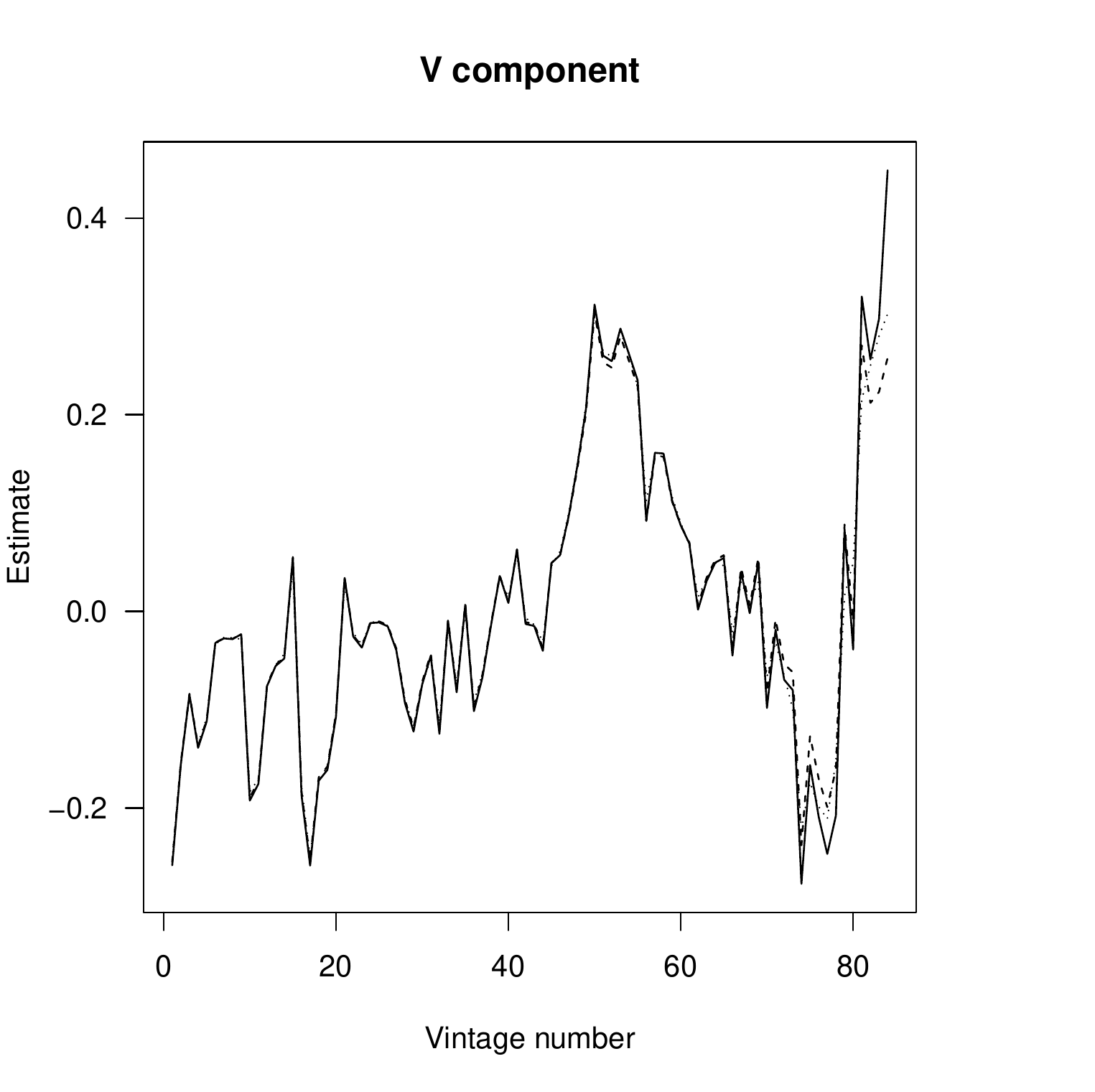}
\caption {Vintage component under fixed effects specification (solid line), independent random effects specification (dashed line) and AR(1) random effects specification (dotted line). }
\label{emvre}
\end{center}
\end{figure}

The top panel represents the vintage parameter estimates $\beta^V$, under the assumption that the vintage parameters are independent and identically normally distributed, that is
$$
\beta_v^V \sim N(0,\sigma_V^2 )
$$
where $\sigma_V^2$ is a vintage variance parameter (estimated here as 0.0155). The lower panel represents the vintage parameter estimates $\beta^V$, under the AR(1) autoregression model, that is
$$
\beta_v^V \sim N(\rho\beta_{v-1}^V,\sigma_V^2 ).
$$
Here, the autoregression coefficient $\rho$ is estimated as 0.822, which confirms the high level of serial dependence observed in the vintage effect. For both random effects models, it can be observed that the random effects estimates are shrunk towards a level which is determined by the other vintages, particularly for the later vintages with fewer observations in the data. This 'borrowing of strength' is exactly what we would expect, and is an attractive feature of random effects models. Overall, however, the degree of shrinkage is not large, as the individual vintage effects are strong. The advantages of the random effects construction will be more evident when some of the between vintage variability is accounted by regressing on vintage-level covariates as suggested above, or for a smaller portfolio.
From a survival analysis perspective, which is relevant when the response variable arises out of a default process, the vintage coefficients can be thought of as representing a shared frailty, a common (to accounts originating in the same time period) tendency to default. In addition to shared frailty, there remains the possibility of unobserved account-level frailty, that is differences between accounts within the same vintage in their tendency to default.

\section{Discussion}

EMV models provide a powerful tool to enhance understanding of the response of a credit portfolio over time. EMV models are linear in structure and of regular form. Standard methods of estimation are generally robust, produce estimators with good properties, and can be flexibly implemented in standard software. 
However, when a nonparametric EMV model is fitted, the decomposition presented is subject to an arbitrary identifiability constraint. In this paper, we have illustrated that care is required in presenting a nonparametric EMV decomposition. We recommend that, whatever software procedure is used to fit the model, the estimates are post-processed to make them consistent with a substantively meaningful identifiability constraint. One possible such constraint (or set of constraints) is based on the limiting behaviour of the maturity curve. 
When a semiparametric model is fitted, using macroeconomic covariates to explain the calendar-time variation in the response, the identifiability issue disappears, though care still needs to be taken if the covariates used are strongly correlated with a time trend, as this may result in a lack of robustness in the estimates.

Further research would be useful in enhancing the utility of EMV models for credit portfolio analysis. One area we have identified, discussed in Section 5.2, is is to investigate the extent to which the maturity effect can be smoothed, either semiparametrically or by modelling it explicitly as a (log) hazard function arising from a given parametric form. The additive form of the EMV model may also be questioned, and the scope for extending the age-time interaction beyond the structured vintage effect investigated.

\bigskip
\parindent 0pt
\hangindent 10pt
\section*{References}

\hangindent 10pt
Aalen, O.O. and Gjessing, H.K. (2001). Understanding the shape of the hazard rate: a process point of view. {\it Statistical Science}, {\bf 16}, 1-22.

\hangindent 10pt
Bosman, M. (2012). {\it The potential of cohort analysis for vintage analysis}. Unpublished M.Sc. thesis. University of Twente, The Netherlands. Available at
{\tt http://essay.utwente.nl/61383/}.

\hangindent 10pt 
Breeden, J.L. (2007). Modeling data with multiple time dimensions {\it Computational Statistics and Data Analysis}, {\bf 51}, 4761-4785.

\hangindent 10pt
Clayton, D., and Schiffers, E. (1987) Models for temporal variation in cancer rates II: age-period-cohort models. {\it Statistics in Medicine}, {\bf 6}, 469-481.

\hangindent 10pt
Fienberg, S.E., and Mason, W.M. (1985). Specification and implementation of age, period and cohort models. In {\it Cohort Analysis in Social Research}, eds. 
W.M. Mason and S.E. Fienberg, 45-88.

\hangindent 10pt
Fu, W.J., Land, K.C. and Yang, Y. (2011). On the intrinsic estimator and constrained estimators in age-period-cohort models.
{\it Sociological Methods and Research}, {\bf 40}, 453-466.

\hangindent 10pt
Glenn, N.D. (1976) Cohort analysts' futile quest: statistical attempts to separate age, period and cohort effects. {\it American Sociological Review}, {\bf  41}, 900-905.

\hangindent 10pt
Goldstein, H.J. (1979) Age, period and cohort effects -- a confounded confusion. {\it Journal of Applied Statistics}, {\bf  6}, 19-24.

\hangindent 10pt
Kupper, L.J., Janis, J.M., Salama, I.A., Yoshizawa, C.N. and Greenberg, B.G. (1985). Age-period-cohort analysis: an illustration of the problems in assessing interaction in one observation per cell data. {\it Communications in Statistics -- Theory and Methods}, {\bf 12}, 2779-2807.

\hangindent 10pt
O'Brien R.M. (2011). Constrained estimators and age-period-cohort models.
{\it Sociological Methods and Research}, {\bf 40}, 419-452.

\hangindent 10pt
Robertson, C., Gandini, S. and Boyle, P.  (1999) Age-period-cohort models: a comparative study of available methodologies. {\it Journal of Clinical Epidemiology}, {\bf  52}, 569-583.

\hangindent 10pt
Siarka, P. (2011). Vintage analysis as a basic tool for monitoring credit risk. {\it Mathematical Economics}, {\bf 7}, 213-228.

\hangindent 10pt
Tu, Y-K, Davey Smith, G. and Gilthorpe, M.S. (2011). A new approach to age-period-cohort analysis using partial least squares regression. {\it PLOS One}, {\bf 6}, e19401, 1-9.

\hangindent 10pt
Therneau, T.M. and Grambsch, P.M. (2000). {\it Modelling Survival data}. Springer, New York.

\hangindent 10pt
Wold, H. (1985). Partial least squares. In {\it Encyclopaedia of Statistical Sciences, Vol 6}, eds. S. Kotz and N.L. Johnson, 581-591.

\hangindent 10pt
Yang, Y., Fu, W.J. and Land, K.C. (2004). A methodological comparison of age-period-cohort models: intrinsic estimator and conventional generalized linear models.
{\it Sociological Methodology}, {\bf 34}, 75-110.

\hangindent 10pt
Zhang, A. (2009). {\it Statistical Methods in Credit Risk Modeling}. Unpublished Ph.D. thesis. University of Michigan, USA.\\ Available at
{\tt http://deepblue.lib.umich.edu/bitstream/handle/2027.42/63707/ajzhang\_1.pdf}.

\end{document}